\documentclass[]{aa}
\begin{document} 
\thesaurus{11(11.13.1; 11.05.2; 09.16.1)} 
\title{Gravitation Wave Emission from Radio Pulsars Revisited}
\author{T. Regimbau, J.A. de Freitas Pacheco} 
\offprints{T. Regimbau}
\institute{Observatoire de la C\^ote d'Azur, B.P. 4229, 06304 Nice Cedex 4, 
France}
\mail{regimbau@obs-nice.fr, pacheco@obs-nice.fr}
\date{Received date; accepted date} 
\maketitle
\markboth{Regimbau.:~GW from Pulsars}{}

\begin{abstract}{We report a new pulsar population synthesis  based on Monte Carlo techniques, aiming to estimate the contribution of galactic radio pulsars
to the continuous gravitational wave emission. Assuming that the rotation periods of pulsars at birth have a Gaussian distribution, we find that the average initial period  is 290 ms. The number of objects with periods equal to or less than 0.4 s, and therefore capable of being detected by an interferometric gravitational antenna like VIRGO, is of the order of 5100-7800. With integration times lasting between  2 and 3 yr, our simulations suggest that about two detections should be possible, if the mean equatorial ellipticity of the pulsars is $\varepsilon$ =10$^{-6}$. A mean ellipticity an order of magnitude higher increases the expected number of detections to 12-18, whereas  for $\varepsilon < 10^{-6}$, no detections are expected}
\end{abstract} 

\section{Introduction}

The first large wide-band gravitational interferometric antennas, the 
French-Italian VIRGO and the American LIGO, should be fully operational at the beginning of the next century. The expected sensitivity of these detectors in terms of the gravitational strain is of the order of h$\approx 10^{-22}$ for short-lived, impulsive bursts, and h$\approx 10^{-26}$ for periodic ''long-lived'' sources for which integration times of about  10$^7$ s are possible. Clearly, a major concern of these detectors is to understand and minimize the sources of noise which define the sensitivity of the antenna, but the identification of a possible ''standard'' source (or sources), having a well defined waveform, is also a relevant problem in the context of the signal  detection.

Among the possible candidates, it is expected that neutron stars, which are 
known to be quite abundant in the Galaxy, could rank aamong conspicuous emitters of gravitational waves (GW). Rotating neutron stars may have a time-varying quadrupole moment and hence radiate GW, by either having a triaxial shape or a misalignment between the symmetry and the spin axes, which produces a wobble in the stellar motion (Ferrari\& Ruffini 1969; Zimmerman \& Szedenits 1979; Shapiro \& Teukolsky 1983; de Ara\'ujo et al. 1994). Moreover, fast rotating proto-neutron stars may develop different instabilities such as the so-called Chandrasekhar-Friedman - Schutz (CFS) instability (Chandrasekhar 1970; Friedman \& Schutz  1978), responsible for the excitation of density waves traveling around the star in the sense opposite to its rotation, or to undergo a transition from axi-symmetric to triaxial shapes through the dynamical ''bar''-mode instability (Lai \& Shapiro 1995). All these mechanisms are potentially able to emit large amounts of energy in the form of GW.

Continuous GW emission by the entire population of rotating neutron stars, 
Assumed to have small deviations from axisymmetry, raises the possibility of detecting their combined signals. The contribution of individual {\it observed} sources, like galactic pulsars, has been considered by Barone et al. (1988), Velloso et al. (1996), among others. However, detected radio pulsars constitute only a small fraction of the actual galactic population. Attempts to estimate the gravitational strain of the total pulsar population have been made by Schutz (1991), Giazotto, Bonazzola \& Gourgoulhon (1997, hereafter GBG97), de Freitas Pacheco \& Horvath (1997), Giamperi (1998). Excepting Schutz (1991), all these authors proposed to detect the {\it square} of the gravitational strain, and the sidereal modulation of the integrated signal was examined by GBG97 as well by Giamperi (1998). In order to perform such an estimation, the actual distribution of the rotation periods must be known. The rotation period is a critical parameter, because the gravitational strain depends on the square of the angular velocity. Unfortunately, the observed distribution does not necessarily reflect the real period distribution due to different selection effects present in all pulsar searches. GBG97 assumed for their estimate an (optimistic) average rotation period of 5 ms, which now seems a rather short value. If we exclude binary millisecond pulsars, which have probably been spun-up by accretion mechanisms, several observational facts indicate that population I pulsars are born with smaller rotation velocities (Bhattacharya 1990). Moreover, several population synthesis calculations also suggest higher initial periods (Narayan 1987; Bhattacharya et al. 1992). Magnetic coupling between the proto-neutron star and the outer envelope could be an efficient mechanism for transferring angular momentum, so reducing the initial angular velocity. Proto-neutron stars are expected to be hot, with temperatures around 10$^9$ K. In this situation, depending on the viscosity, a rapidly rotating proto-neutron star may excite r-modes, emitting GW which decelerates considerably the star. According to computations by Andersson, Kokkotas \& Schutz (1999), this mechanism sets a limit around P $\approx$ 20 ms for the fastest newly born neutron stars. However, it should be emphasized that this value is a function of the adopted (uncertain) viscosity law. 

In the present work we revisit the problem of the contribution to the continuous GW emission from the pulsar population inside the galactic disk. We estimate the  actual period distribution by using population synthesis based on Monte Carlo methods, looking for the best pulsar population parameters by comparing our simulation results with data. Using the model population, we compute the gravitational strain for each pulsar emitting GW in the frequency  band of VIRGO, the total square of the signal as well the amplitude modulation due to the sidereal motion. We show, in agreement with the expectations made by de Freitas Pacheco \& Horvath (1997), that the signal is strongly dominated by a few fast pulsars and/or by the nearest ones. The plan of this paper is the following: in section 2 we describe our population synthesis method; in section 3 we present the properties of the simulated population; in section 4 we estimate the contribution to the continuous gravitational strain of such a population and finally, in section 5 we present our main conclusions. 

\section{The Population Synthesis}

\subsection{The Rotation Period Evolution}

The evolution of the rotation period of a neutron star, in the framework of the magnetic dipole model, is given by (Pacini 1968) 
\begin{equation}
{P\dot P} = {{8\pi^2}\over{3}}{{\mu^2}\over{Ic^3}}sin^2\alpha
\end{equation}
where $\mu$ = ${{1}\over{2}}BR_p^3$ is the surface magnetic moment, B is the magnetic field, R$_p$ and I are respectively the radius and the moment of inertia of the neutron star, c is the velocity of light and $\alpha$ is the angle between the rotation axis and the magnetic dipole. Integration of this equation with constant $\alpha$ gives the rotation period evolution of the so-called standard model.

In the absence of abrupt changes of the magnetic torque or of the internal structure of the star, solutions of eq. (1) indicates a smooth increase of the rotation period with time. However, most pulsars do not slow down regularly, but display variations in their spin rates in the form of glitches (Shemar \& Lyne 1996) and timing noise. An important aspect of glitches is the persistent increase of the spin-down rate following these events, which could be due to a sudden and permanent increase of the external torque (Link \& Epstein 1997).In order to explain the glitches observed in the spin evolution of the Crab pulsar, Alpar \& Pines (1993) suggested a reduction of the moment of inertia, induced by changes in the internal structure of the star. This would produce a spin-up of the star due to angular momentum conservation, but a decrease in the spin-down
rate, contrary to observations. As the star spins down, it becomes less oblate, inducing stresses that may lead to starquakes if the yield strength of the solid crust is exceeded. Starquakes may affect the braking mechanism if they are able to change the position of the magnetic moment with respect to the rotation axis. If the structural relaxation after a starquake occurs asymmetrically about the 
rotation axis due, for instance, to magnetic stresses, the figure and spin axes may become misaligned. Under this condition, the star precesses and relaxes to a new equilibrium state, corresponding to a new orientation of the magnetic dipole with respect to the rotation axis. This possibility was examined in a recent work by Link, Franco \& Epstein (1998). In their picture, a spin-down pulsar reduces its equatorial radius by shearing material across the equator and moving material along faults to higher latitudes. Such  crustal motions may produce an increasing misalignment between the rotation and magnetic axes, providing a natural explanation for the observed increases in the spin-down rate following glitches in the Crab, PSR 1830-08 and PSR 0355+54. 

The migration of the magnetic dipole fromm an initial angle $\alpha$ to an orthogonal position with respect to the spin axis was also considered by Allen \& Horvath (1997). They showed that with such a variable torque, non-canonical braking indices may be explained. Here we adopt the scenario developed by Link, Franco \& Epstein (1998), assuming that the magnetic dipole migrates from an arbitrary initial angle to a final orthogonal position in a timescale t$_{\alpha}$. 
For our present purposes, this migration is modeled by the equation
\begin{equation}
sin^2(\alpha (t)) = 1 - n_0e^{-t/t_{\alpha}}
\end{equation}
We emphasize that this equation is not the result of any physical model, but only a phenomenological representation of the above picture, in the sense that a given pulsar born with an initial angle $\alpha_0 = acos(\surd n_0)$, will reach an orthogonal configuration in a timescale $t_{\alpha}$.

Combining eq. (1) and (2) and integrating, we obtain for the evolution of the rotation period
\begin{equation}
P = P_0\lbrack 1 + {{t}\over{\tau_0}} - n_0{{t_{\alpha}}\over{\tau_0}}(1 -
e^{-t/t_{\alpha}})\rbrack^{1/2}
\end{equation}
where P$_0$ is the initial rotation period and $\tau_0$ = ${{3}\over{4\pi^2}}{{Ic^3P_0^2}\over{B^2R^6}}$ is the magnetic braking timescale. For $t >> t_{\alpha}$ we recover essentially the standard model evolution. However,  eq. (3) allows one to explain the existence of objects with short periods and with a relatively small deceleration rate, which would be the consequence of a small initial misalignment between the magnetic and the spin axes, but not due to a small magnetic field strength. Note that in our picture these pulsars are young, in spite of having a large indicative age t$_s$ = ${{P}\over{2\dot P}}$.

Another important parameter characterizing the evolution of the rotation period is the so-called braking index $N$ defined as
\begin{equation}
N = {{\ddot\Omega\Omega}\over{\dot\Omega^2}}
\end{equation}

For the standard magnetic dipole model, the expected braking index is $N$ = 3.
However, the few pulsars with $N$ measured accurately show deviations from the 
Standard model; $N$=2.518 for the Crab, $N$=2.837 for PSR1509-58, $N$=2.04 for 
PSR0540-69. The exceptionally low value for the Vela pulsar $N$ = 1.40 was recently reported, suggesting that the braking torque is not due to a ''pure'' stationary dipole field (de Freitas Pacheco \& Horvath 1997). From eqs. (3) and (4), the predicted braking index for our model is
\begin{equation}
N =  \lbrack 3 - 2({{t_s}\over{t_{\alpha}}})ctg^2(\alpha(t)) \rbrack
\end{equation}
This equation indicates that for young pulsars with ages less than $t_{\alpha}$ the braking index $N$ may be less than three, tending to the canonical value when $t >> t_{\alpha}$.

The angle $\alpha$ may vary ''discontinuously'' when the crust suddenly cracks
or almost ''continuously'' as eq. (2) suggest. It is worth mentioning that
pulsars like PSR1509-58 and PSR0540-69  have not yet displayed any glitch
activity even though their braking indices are less than the canonical value $N$ 
= 3. This could be an indication that  other braking mechanisms are operative such as, for instance, anomalous electro-magnetic torques recently discussed by
Casini \& Montemayor (1998).

For the Crab pulsar, since the true age, the braking index $N$ and $t_s$ are known, we can compute using the equations above the dipole migration rate $<ctg\alpha\cdot{{d\alpha}\over{dt}}>$, which is about 9.2$\times 10^{-5}$ rad/yr  for a solution where the present angle between the spin and the dipole axes is about 64$^o$. From glitch data spanning a time interval of about 20 yr, Allen \& Horvath (1997) find a migration rate 1.8$\times 10^{-5}$ rad/yr, which is about five times smaller than the predicted value. However, as those authors 
remarked, such a time interval is still very short and discrepancies should be expected when a comparison is performed between the migration rate derived from the average of few events and that obtained from a continuous variation model.
For the other remaining three pulsars, the average ''continuous'' migration rate 
is $ <ctg\alpha\cdot{{d\alpha}\over{dt}}>$ = 6.5$\times 10^{-5}$ rad/yr (Allen \& Horvath 1997). These rates are comparable to within an order of magnitude, reinforcing the idea of a common origin and giving some support to our adopted spin-down model. 

\subsection{The Method}

Initial spin periods and magnetic fields are important parameters tracing the
formation history of neutron stars and their interaction with the surrounding medium. The pulsar population study by Narayan (1987) raised the so-called ''injection'' problem, which would connect the initial period and magnetic field in the sense that slow born neutron stars would be associated with high field strengths. Unfortunately,  the ages of most  pulsars are unknown and present measurements for estimating the period and  its time derivative are insufficient in order to estimate the initial field and spin period. In this case, methods of population synthesis can be useful tools to estimate statistical values of these parameters, assuming ad-hoc distribution laws. Moreover, the {\it observed} pulsar population is strongly dominated by luminous objects, and since the radio emission depends on the spin period as well as on the spin down rate, present samples are biased, excluding long-period and high-field objects as well as old and faint pulsars. Comparison with actual data requires the inclusion of these selection and observational bias in the generation of synthetic catalogs, which are more treatable when Monte Carlo techniques are employed.

In order to generate a synthetic sample of pulsars, we need to model: (a) their 
Distribution in the Galaxy, (b) the initial parameters affecting their evolution, (c) the effects of the interstellar medium in their detection and, (d)  the sensitivity of a given survey.

The distribution in the Galaxy was modeled by assuming a constant birthrate until a maximum age t$_{max}$, considered as a free parameter in our numerical experiments. The ratio between the total number N$_p$ of ggenerate pulsars and t$_{max}$ gives the mean birthrate. For a given experiment, N$_p$ is the number of input pulsars required to reproduce the  number of objects in the data sample, when observational and selection constraints are imposed.
 
Pulsars are strongly concentrated in the galactic plane and their distribution along the z-axis can be represented by an exponential law with a scale of height of about 100 pc. Some previous studies suggest a possible correlation between the height above the galactic plane and age or magnetic field (Narayan \& Ostriker 1990), but here we assume that the initial z-coordinate is a statistically independent variable. In the galactic plane, the density of pulsars probably decays exponentially, in the same way as population I objects.
Recent galactic mass profile models, based on infrared data, suggest a short distance scale of about 2.3 kpc (Ruphy et al. 1996); this was adopted in our numerical experiments. In order to compute heliocentric distances, the galactocentric distance of the sun was assumed to be 8.5 kpc. Some pulsars have significant proper motions, implying large initial velocities that must be taken into account when computing distances, since they affect the orbital
motion in the Galaxy. An isotropic Gaussian velocity distribution  with 1-D  velocity dispersion of  100 km/s  (Lorimer et al. 1993) was adopted to simulate 
the pulsar kick velocity, but our results are not significantly changed if the velocity dispersion is increased up to 200 km/s. Once the initial R and z coordinates for a given pulsar were established, if its kick velocity is zero, we assumed that the pulsar follows a circular orbit using the galactic rotation  curve by Schmidt (1985). For non-zero kick velocity, the new energy and angular momentum of the orbit were computed and the object was settled in a new trajectory. The heliocentric distance is computed after a time t (the age of the pulsar), taking into account the differential rotation between the sun and the
pulsar. The present galactic coordinates ( l, b) of the object are also computed, since these are required to establish if the pulsar should be included or not in the area of the sky covered by  a given survey, as we shall see later.

In the present simulations we have assumed that the initial rotation period
of pulsars can be represented by a Gaussian distribution, characterized by a mean value P$_0$ and a dispersion $\sigma_{P_0}$. We have assumed that the break-up velocity is about 12000 rad/s, thus pulsars generated with  rotation periods less than 0.5 ms were not included in the simulated sample. These parameters were allowed to vary in order to verify their effects on the results of different numerical experiments. A similar procedure was adopted for the parameter $\tau_0$ (see, eq. 3), which is related to the magnetic field of the pulsar, for which a log-normal distribution was assumed. Simulations by Bhattacharya et al. (1992) and  by Mukherjee \& Kembhavi (1997) suggest that the timescale for magnetic field decay is longer then the pulsar lifetime ($\tau_D >$ 100-160 Myr), thus we have assumed a constant field in our numerical experiments. The initial angle $\alpha_0$ (or, equivalently, the value of n$_0$)
between the magnetic dipole and the spin axes is calculated supposing a random distribution. Once these parameters are assigned for a given pulsar ($t_{\alpha}$ is kept fixed during a giving experiment), the period and the time derivative can be computed from eq.(3).

The comparison between a simulated sample with actual data requires the evaluation of the expected radio flux density S$_{\nu}$ at a given frequency (400 Mhz in general). To compute S$_{\nu}$, one needs to know the distance and the luminosity of the pulsar. A theory able to predict the pulsar radio luminosity is still missing, but the spin period and its time derivative are expected to be the main variables on which the radio power depends, since the magnetic field can be expressed in terms of these variables in the standard model. Different power laws have been suggested in the literature connecting $P$ and $\dot P$ (Lyne, Ritchings \& Smith 1975; Stollman 1987; Emmering \& Chevalier 1989).  When these laws are compared with observational data a large dispersion is found, which can be explained by a bad understanding of the radio emission mechanism and/or by errors in the distance, estimates of which are derived by using the dispersion of radio waves through the interstellar plasma and by adopting a model for the electron density distribution inside the galactic disk. The effects of the adopted  luminosity law in numerical
simulations were discussed by Lorimer et al. (1993), who concluded that, at the
present time there appears to be no satisfactory model to account for observed pulsar luminosities.

The radio luminosity ought to be a decreasing function of the spin period, since
slow pulsars missing in different surveys are probably too weak to be detected. Models assuming a radio luminosity proportional to the cube root of the spin-down rate lead to an inverse dependence on the period (Narayan 1987), while
other proposed power laws indicate even higher exponents. In the present work, 
we have adopted a different empirical luminosity law, imposing the following 
exponential fate on the radio emission
\begin{equation}
L_r = A{\dot P^{\alpha}}e^{-\gamma P}
\end{equation}
The parameters in this relation were determined by an optimised fitting procedure, using the upgraded pulsar catalog of Taylor, Manchester \& Lyne (1993). If the luminosity is given in mJy.kpc$^2$, and the period is in seconds, the resulting numerical values of the parameters are: A=3.3$\times 10^7$, $\alpha$ = 0.34 and $\gamma$ = 1.02. These figures are slightly different from
those given by de Freitas Pacheco \& Horvath (1997) because here we have used
a larger sample in the fitting procedure. 

The radio emission is not isotropic, implying that pulsars are not visible from all directions, and this imposes another constraint on their detectability.  Following the approach used by Emmering \& Chevalier (1989), which will be adopted here, the beaming fraction of the sky covered by the pulsar emission cone is
\begin{equation}
f = (1 - {\rm \cos\theta}) + ({{\pi}\over{2}} - \theta){\rm \sin\theta}
\end{equation}
where $2\theta$ is the aperture of the emission cone. Since short period pulsars seem to have wider emission cones, we have adopted the beaming model proposed by Biggs (1990), which gives for the cone aperture
\begin{equation}
\theta \approx {{6.2}\over{\surd P}}
\end{equation}
where $\theta$ is in degrees and $P$ is in seconds. It is worth mentioning that the study by Rankin (1990) of a large sample of pulsars also indicates a variation of $\theta$ inversely proportional to the square root of the period.

Our data are a culled sample of 389 pulsars extracted from the catalog of Taylor, Manchester \& Lyne (1993), where only objects that have originated from population I stars and with all required parameters measured were considered. Pulsars not included in any of the original surveys were also discarded. Since this catalog is a compilation of different catalogs covering specific sky areas and having a well defined sensitivity, the objects were distributed in sub-classes, according to the original surveys in which they were first detected.

Each simulated catalog to be compared with these data is prepared with the
following prescriptions. For each pulsar generated with a given set of initial
parameters, we have followed its evolution in order to compute the present $P$, $\dot P$ values and  galactic coordinates, according to the scheme described above. Then the radio luminosity and the flux density  $S_{\nu}$ = ${{L_r}\over{d^2}}$ are calculated, as well as the probability for the pulsar beam to cross the earth. The galactic coordinates define the survey (or surveys) included in the general catalog  that covered this region. Then, we compare the expected flux density with the minimum detectable flux density of that survey through the relation (Dewey et al. 1984; Manchester et al. 1996)
\begin{equation}
S_{min}=A_0(T_r + T_{sky})\lbrack{{W}\over{(P-W)}}\rbrack^{1/2}
\end{equation}
where W is the observed (broadened) pulse width, T$_r$ and T$_{sky}$ are respectively the system and the sky background temperature in the considered line of sight. A$_0$ is a normalization constant, defined as
\begin{equation}
A_0 = {{(S/N)\beta}\over{G_a\surd (n_p\Delta \nu t_{int})}}
\end{equation}
In this equation, G$_a$ is the antenna gain, n$_p$ is the number of polarizations used, $\Delta\nu$ is the receiver bandwidth, t$_{int}$ is the integration time, S/N is the signal-to-noise ratio and $\beta$ is a numerical factor, including  other loss processes. The variation of the sky temperature with the galactic coordinates was estimated using the relation by Narayan (1987). The intrinsic pulse width W$_e$ was assumed to be equal to 4\% of the period ( W$_e$=0.04P) and to model the pulse broadening due to instrumental effects, dispersion and scattering, we follow the approach by Dewey et al.(1984) (see also Lorimer et al. 1993), writing for the relation between the observed and the intrinsic pulse width
\begin{equation}
W^2 = W_e^2 + \tau^2_{sam}+\tau^2_{sca}+\tau^2_{DM}
\end{equation}
where $\tau_{sam}$ is the effective sampling time, $\tau_{sca}$ is the broadening due to diffusion through the interstellar medium and $\tau_{DM}$ is the channel broadening. We have used the relation given by Bhattacharya et al. (1992) for $\tau_{sca}$, namely,
\begin{equation}
\tau_{sca}(ms) =10^{-4.62+1.14log(DM)}+10^{-9.22+4.46log(DM)}
\end{equation}
where DM is the electron dispersion measure in the line of sight of the pulsar 
in pc.cm$^{-3}$. For the channel broadening, we have used the relation 
\begin{equation}
\tau_{DM}(\mu s) = C_{DM}.DM 
\end{equation}
where, following Stokes et al. (1986)
\begin{equation}
C_{DM} = 8.3{{B_{ch}(MHz)}\over{\nu^3(GHz)}}
\end{equation}
with B$_{ch}$ being the channel bandwidth and $\nu$  the observing frequency. Table 1 gives the main parameters characterizing the different catalogs adopted
in our simulations. Note that values inside braces correspond to high latitude
data.   

\begin{table*}
\caption[1]{Survey Parameters}
\begin{flushleft}
\begin{tabular}{lcccccccccc}
\noalign{\smallskip}
\hline
\noalign{\smallskip}
Survey& T$_r$ (K)& $\tau_{sam}$ (ms)& C$_{DM}$(ms.pc$^{-1}$.cm$^3$)& A$_0$(mJy/K)\\
\noalign{\smallskip}
\hline
\noalign{\smallskip}
Arecibo 1& 110 & 33 & 0.0094 & 0.038\\
Arecibo 2& 90 & 0.6 & 0.0063 & 0.085\\
Arecibo 3-4 & 62 & 1.0(0.5) & 0.0082(0.026)& 0.02(0.032)\\
Green Bank 1 & 170 & 33 & 0.28 & 0.30\\
Green Bank 2 & 30 & 33 & 0.28 & 0.13\\
Green Bank 3 & 30 & 4.0 & 0.035 & 0.18 \\
Jodrell Bank & 110 & 80 & 0.30 & 0.36 \\
Molonglo 2 & 210 & 40 & 0.06(0.3)& 0.17 \\
Parkes & 30 & 0.60 & 0.0125 & 0.19 \\
\noalign{\smallskip}
\hline
\end{tabular}
\end{flushleft}
\end{table*}

Our procedure insures that the simulated catalog will have the objects distributed in the same proportion as in the global catalog, taking into account the specific sensibility and the sky area covered by each survey. The numerical experiments generate objects according to these prescriptions until the simulated number of  detectable pulsars is equal to the number in our data sample. For a given experiment, defined by a set of initial parameters, the number of runs is comparable to the number of objects in the sample and the final result is a suitable average, in order to avoid statistical fluctuations. 

For each experiment so defined, we have compared the resulting  distributions with those derived from actual data. We searched for optimize the various input parameters that describe the pulsar population, controlling the fit quality through $\chi^2$ and  Kolmogorov-Smirnov tests.

\section{Properties of the Simulated Pulsar Population}

The possibility that the magnetic axis could migrate gives not only a possible 
explanation for the non-canonical braking index observed in some pulsars, but 
also a mechanism affecting the $\dot P$ and age distributions. In general, objects with small deceleration rates ($\dot P \approx 10^{-16}$) are old and have low magnetic fields in the standard picture. In our scenario the magnetic torque controlling the deceleration rate, depends not only on the field strength but also on the angle between the dipole and the spin axes. Therefore, our simulated catalog contains some young objects born with a small misalignment between those axes, having a high field strength but a small deceleration rate.  Unfortunately, it is not possible to disentangle easily both situations and, as a consequence, the derived value of the dipole migration timescale t$_{\alpha}$
has a large uncertainty: $t_{\alpha}$ = 10000$\pm$4000 yr. These values imply migration rates spanning the interval (2 - 18)$\times 10^{-5}$ rad/yr, which compare with  rates derived by other authors (see section 2.1). 

The optimized parameters characterizing the distribution of the initial period and that of the magnetic braking timescale $\tau_0$ are given in table 2 for t$_{max}$ equal  to 24 Myr and 100 Myr. These parameters represent the best compromise able to fit adequately to data, the simulated distributions of period and its derivative, as well as the distribution of distances to the sun. Note that the parameters characterizing the Gaussian representing the initial distribution of rotation periods are not sensitive to the adopted value for $t_{max}$, but this is not the case for the parameter $\tau_0$.

In figure 1 we have plotted in the plane XY, which coincides with the galactic plane (the origin of the coordinates is placed at the galactic center), the observed pulsar population and our simulated population. The superposition of both populations is quite good, suggesting that selection effects were adequately taken into account by the model. In figure 2 we compare the observed period distribution with simulated data, where error bars correspond to the dispersion derived from five hundred experiments. Figure 3 shows the comparison
between observed and simulated data for the period derivative, whereas figure 4
display both heliocentric distance distributions. These figures correspond to
simulations with $t_{max}$ = 24 Myr and they illustrate the fit quality.

\begin{table*}
\caption[2]{Pulsar Population Properties}
\begin{flushleft}
\begin{tabular}{lcccccc}
\noalign{\smallskip}
\hline
\noalign{\smallskip}
Parameters& t$_{max}$ = 24 Myr& t$_{max}$ = 100 Myr \\
\noalign{\smallskip}
\hline
\noalign{\smallskip}
P$_0$ (ms) & 290$\pm$20 & 290$\pm$20\\
$\sigma_{P_0}$ (ms) & 90$\pm$20 & 90$\pm$20\\
ln($\tau_0$) (yr) & 9.0$\pm$0.5 & 11.0$\pm$0.5\\
$\sigma_{ln(\tau_0)}$ (yr) & 3.6$\pm$0.2 & 3.4$\pm$0.2 \\
$<logB>$ (G)& 12.4 & 12.1\\
\noalign{\smallskip}
\hline
\end{tabular}
\end{flushleft}
\end{table*}

The last line of table 2 gives the average magnetic field of the simulated observed population, which is about 2.5$\times 10^{12}$ G, in agreement with values derived from a direct application of the ''standard'' pulsar model. However, the average field of the true or ''unseen'' population is one order of magnitude higher, namely, 2.5$\times 10^{13}$  G. Most of these high-field pulsar have rather long periods and thus are radio-quiet. The relevance of this high-field pulsar population in the context of magnetars will be discussed in a forthcoming paper (Regimbau \& de Freitas Pacheco 2000).

Our simulations indicate that the data are better explained if an initial 
distribution of periods is assumed instead of an unique (mean) initial value. Bhattacharya et al. (1992) assumed an initial period equal to 100 ms, comparable with the average value derived from our simulations (290 ms). Models by Lorimer et al. (1993) are not directly comparable, since they have assumed a magnetic field decay with a timescale of  10 Myr. For the present purposes, we will focus our attention on the pulsar population with periods less than 0.4 s, which contributes to the continuum GW emission from the galactic disc. This upper bound is expected to be attained in a second phase of the VIRGO experiment.
In spite of the bulk of the population having higher periods, the dispersion in 
The initial periods guarantees the existence of objects in that frequency band. From our simulations, we predict about 60-90 (single) pulsars with  P $<$ 80 ms in the Galaxy, a number not in contradiction with present data. The period distribution for pulsars satisfying the condition P $<$ 0.4 s is shown in figure 5, and their heliocentric distance distribution is given in figure 6. 
The number of this sub-population is in the range 5100 - 7800, according to our simulations performed with t$_{max}$ equal to 24 Myr and 100 Myr respectively. The contribution of this population to the gravitational strain is estimated in the next section. 

\section{The Gravitational Strain}

\subsection{The Equations}

As we have mentioned above, pulsars could emit GW by having a time-varying quadrupole moment produced either by a slight asymmetry in the equatorial plane ( assumed to be orthogonal to the spin axis) or by a misalignment between the symmetry and angular momentum axes, case in which a wobble is induced in the star motion. In the former situation the GW frequency is equal to twice 
the rotation frequency, whereas in the latter two modes are possible: one in 
which the GW have the same frequency as the rotation, and another in which the GW have twice the rotation frequency. The first mode dominates by far at small wobble angles while the importance of the second increases for large values.

Here we neglect the possible precessional motion and, in this case, the two polarization components of GW emitted by a rotating neutron star are (Zimmerman \& Szedenits 1979; Bonazzola \& Gourgoulhon 1996 )
\begin{equation}
h_+(t)  = 2A(1+\cos^2i)\cos(2\Omega t)
\end{equation}
and
\begin{equation}
h_{\times}(t) = 4Acosi.sin(2\Omega t)
\end{equation}
where {\it i} is the angle between the spin axis and the wave propagation vector, assumed to coincide with the line of sight, 
\begin{equation}
A = {{G}\over{rc^4}}\varepsilon I_{zz}\Omega^2
\end{equation}
G is the gravitation constant, c is the velocity of light, r is the distance to 
The source, $\Omega$ is the angular rotation velocity of the pulsar and the
ellipticity $\varepsilon$ is defined as
\begin{equation}
{\varepsilon} = {{I_{xx}-I_{yy}}\over{I_{zz}}}
\end{equation}
with the $I_ {ij}$ being the principal moments of inertia of the star.

The detected signal by an interferometric antenna is
\begin{equation}
h(t) = h_+(t)F_+(\theta ,\phi ,\psi) + h_{\times}(t)F_{\times}(\theta ,\phi 
,\psi)
\end{equation}
where $F_+$ and $F_{\times}$ are the beam factors of the interferometer, which 
are functions of the zenith distance $\theta$, the azimuth $\phi$ as well as of 
the wave polarization plane orientation  $\psi$. Notice that the angles $\theta$ 
and $\phi$ are functions of time due to the Earth's rotation, introducing a 
modulation of the signal. The explicit functions for the beam factors, taking into account the geographic localization and the orientation of the VIRGO antenna were taken from Jaranowski, Kr\'olak \& Schutz (1998).

Concerning the detection strategy, recall that the population derived from our simulations of potential GW emitters is in the range 5100 - 7800. In this case, according to the conclusions by GBG97, it becomes advantageous to search for individual detections  instead of the total square amplitude. Here we have simulated both strategies. In the former case, the strain amplitude was calculated for each pulsar satisfying the condition P $<$ 0.4s, using eq.(19) and assuming a random orientation for the inclination {\it i} of the spin axis as well for the orientation of the polarization angle $\psi$. Since the signal is modulated at twice the rotation frequency, in general much shorter than the sidereal period, we have assumed also a random phase for the relative orientation of the detector with respect to the equatorial coordinate system. In the other case, the procedure adopted to calculate the total square of the strain was the following: first, we have squared equ.(19) and then performed an 
average in the time interval $\tau$, satisfying the condition, 0.4 s $ < \tau < $ one day. In this case, the cross terms give a null contribution and we obtain

\begin{equation}
<h^2(t)> = A^2\lbrack 2(1+\cos^2i)^2 F_+^2(t) + 8\cos^2iF_{\times}^2(t) \rbrack
\end{equation}

This equation was used  to compute the contribution from all simulated objects satisfying P $<$ 0.4 s, assuming again a random orientation for {\it i} and $\psi$. 

\subsection{The Results}

The main differences between the present approach and previous calculations should be emphasized. Our procedure allows a more realistic estimate of the rotation period distribution, as well the number of pulsars able to contribute to the gravitational strain. Additionally the spatial distribution of those pulsars throughout the galactic disc changes for each simulation, although their average properties remaining constant. It is thus preferable to present the statistics of our numerical experiments, from which is possible to estimate the
probability of having a signal above a given threshold.

In figure 7 we give the statistics for the gravitational strain {\it h}. For 
each experiment we have computed the distribution of values of  {\it h} and then averaged the results of 500 experiments. Error bars indicate the rmsd for each bin of thickness $\Delta log(\mid h \mid)$ = 0.20. The resulting distribution can be represented quite well by a Laplace law, namely
\begin{equation}
L(x) = e^{(-2\mid x+29.7\mid)}
\end{equation}
where x = log$(\mid h\mid)$. This function corresponds to calculations performed 
with an ellipticity $\varepsilon$ = 10$^{-6}$. For other values of $\varepsilon$, it is enough to replace the constant in eq.(21) by 23.7 log($\varepsilon$). The probability per pulsar tohave a signal above x$_0$ is
\begin{equation}
p(x_0) = \int^{\infty}_{x_0}L(x)dx
\end{equation} 
For a continuous source, VIRGO (or LIGO) will be able to detect amplitudes of the order of {\it h}$\approx 10^{-26}$, with integration times of about 2-3 yr.
From the equation above and the population number estimated previously, one should expect to detect about 2 objects if $\varepsilon = 10^{-6}$ or about 12-18 pulsars if $\varepsilon = 10^{-5}$. If the average ellipticity is smaller
than 10$^{-6}$, most of the objects will be below the detection threshold, at 
least for the present antenna sensibility.

In figure 8 we present the statistics for the square of the amplitude, resulting from the different spatial distribution of the pulsars in each numerical experiment. 
We emphasize that the amplitude of the signal is due essentially to a few pulsars, in agreement with the conclusions derived from the statistics of single
objects and with the analytical study by de Freitas Pacheco \& Horvath (1997). As a consequence, the sidereal modulation is not fixed by the galactic center-anticenter asymmetry, but by those few dominant objects. No typical modulation curve was obtained, since the relative positions of these pulsars vary from experiment to experiment. Using the equations by GBG97 for the signal-to-noise ratio, one should expect to detect a signal of $<h^2> \approx 10^{-45}$ with the presently planned VIRGO sensibility. Our simulations indicate that, if $\varepsilon = 10^{-6}$ a signal of such an amplitude has weak probability to be detected (about 1/100) and signals of the required amplitude can only be
obtained if the average pulsar ellipticity is of the order of  10$^{-4}$.

\section{Conclusions}

Estimates of the GW emission from the total pulsar population in the Galaxy require knowledge of the period and distance distributions. Selection and detection bias are strong and population synthesis methods are necessary in order to model these effects,to recover the ''true'' population properties from the observed distributions.

Pulsars are probably born with a range of periods and magnetic fields. Here we
have assumed {\it ad-hoc} initial distributions for these parameters, which were
optimized in order to reproduce actual data. Our numerical experiments suggest 
that the mean initial period at the pulsar birth is equal to 290 ms, confirming 
the conclusions of other simulations. Nevertheless, we predict about 60-90 single pulsars in the galaxy with periods less than 80 ms, as a consequence of the dispersion in the initial periods. 

The pulsar population satisfying the condition P$<$0.4 s, able to contribute to 
the GW emission within the pass-band of VIRGO, amounts to about 5100-7800 objects. These numbers are considerably smaller than the estimate adopted by GBG97 (1.4$\times10^5$). Those authors assumed that the population with P$<$ 0.4 corresponds to about 28\% of the total population. The adopted fraction is that derived from cataloged pulsars. However, our simulations indicate that this fraction is only about 3.5\%, exemplifying the consequences of the detection bias present in all surveys. 

VIRGO will be able to detect a gravitational strain amplitude of h$\approx 10^{-26}$  with three years of integration. Under these conditions, our simulations indicate 2 detections if $\varepsilon = 10^{-6}$ and up to 12-18 detections if  $\varepsilon = 10^{-5}$. If the average ellipticity is smaller than 10$^{-6}$, no detections are expected, at least with the presently planned antenna sensitivity. The total square amplitude resulting from this population is below the VIRGO threshold and a detectable signal would be produced only if the average ellipticity is at least of order of $10^{-4}$, a value which seems to be excluded by a recent re-analysis of the magnetic and gravitational torques in some (observed) young pulsars (Palomba 1999).

We emphasize that the present results concern uniquely the radio pulsar population and the contribution of a binary millisecond population to the continuous GW emission remains to be estimated. These simulations are more difficult,since the rotation period evolution has a complex history, including 
decelerating and accelerating phases, which will be discussed in a future paper.

\newpage

\noindent
{\bf Figure Captions}
\vskip0.5cm

\noindent
{\bf Figure 1}

\noindent
Spatial distribution in the galactic plane (origin at the galactic
center) of both observed (stars) and simulated (circles) populations
of radio pulsars. The position of the sun is also indicated in the
diagram. 

\noindent
{\bf Figure 2}

\noindent
Distribution of rotation periods. Bars represent observed binned data and filled 
circles
indicate the result of our simulations. Error bars give the rmsd after averaging 
500 numerical
experiments.

\noindent
{\bf Figure 3}

\noindent
Distribution of period derivatives. Symbols have the same meaning as in figure 
1.

\noindent
{\bf Figure 4}

\noindent
Distribution of heliocentric distances. Symbols have the same meaning as in 
figure 1.

\noindent
{\bf Figure 5}

\noindent
Distribution of periods for all pulsars satisfying P$<$0.4 s.

\noindent
{\bf Figure 6}

\noindent
Distribution of heliocentric distances for the pulsar population satisfying 
P$<$0.4 s.

\noindent
{\bf Figure 7}

\noindent
Distribution of the gravitational strain h for $\varepsilon = 10^{-6}$. Error 
bars indicate
rmsd after averaging 500 numerical experiments.

\noindent
{\bf Figure 8}

\noindent
Distribution of the total h$^2$ for $\varepsilon = 10^{-6}$, corresponding to 
7000 numerical
experiments. 

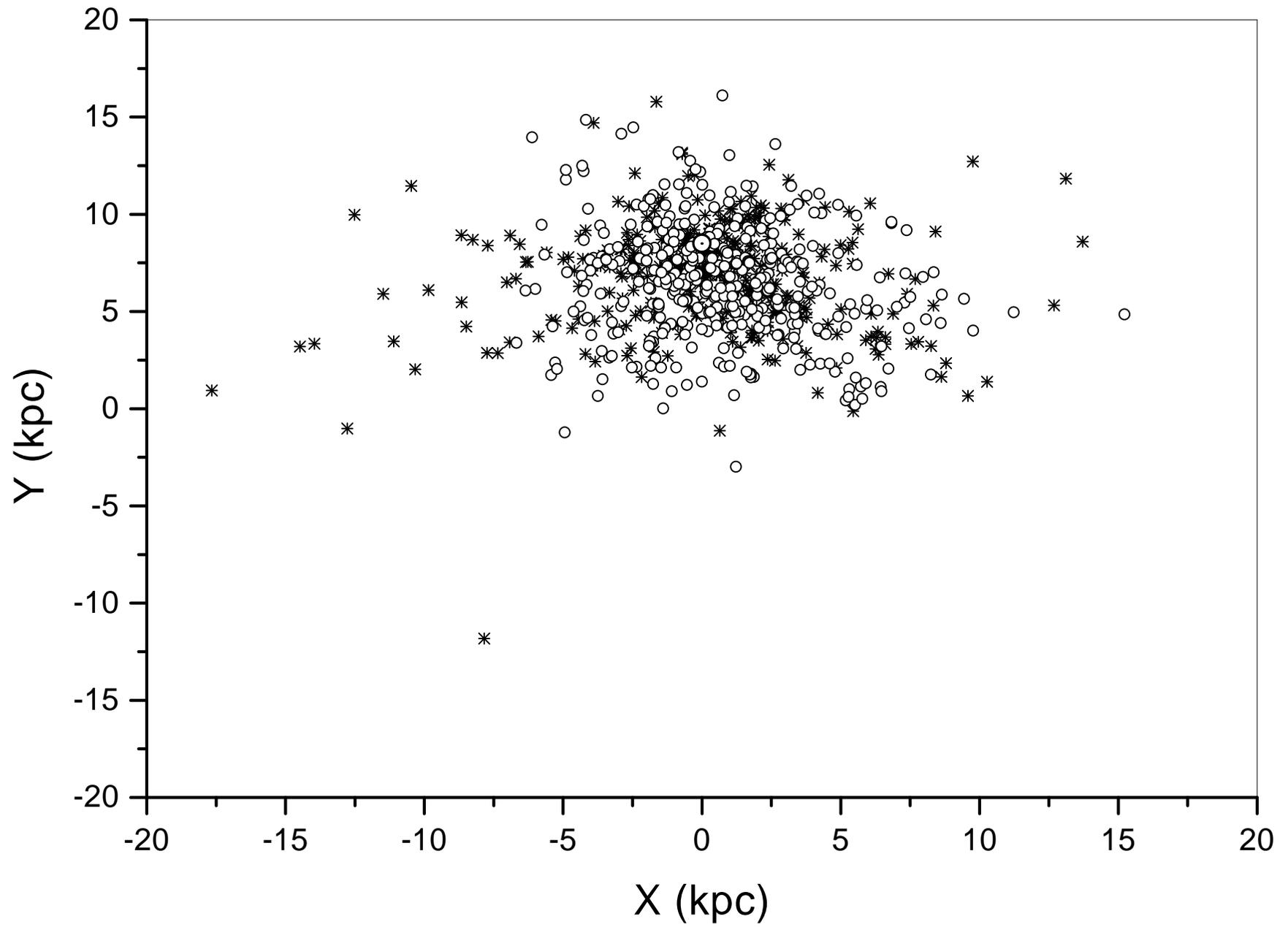
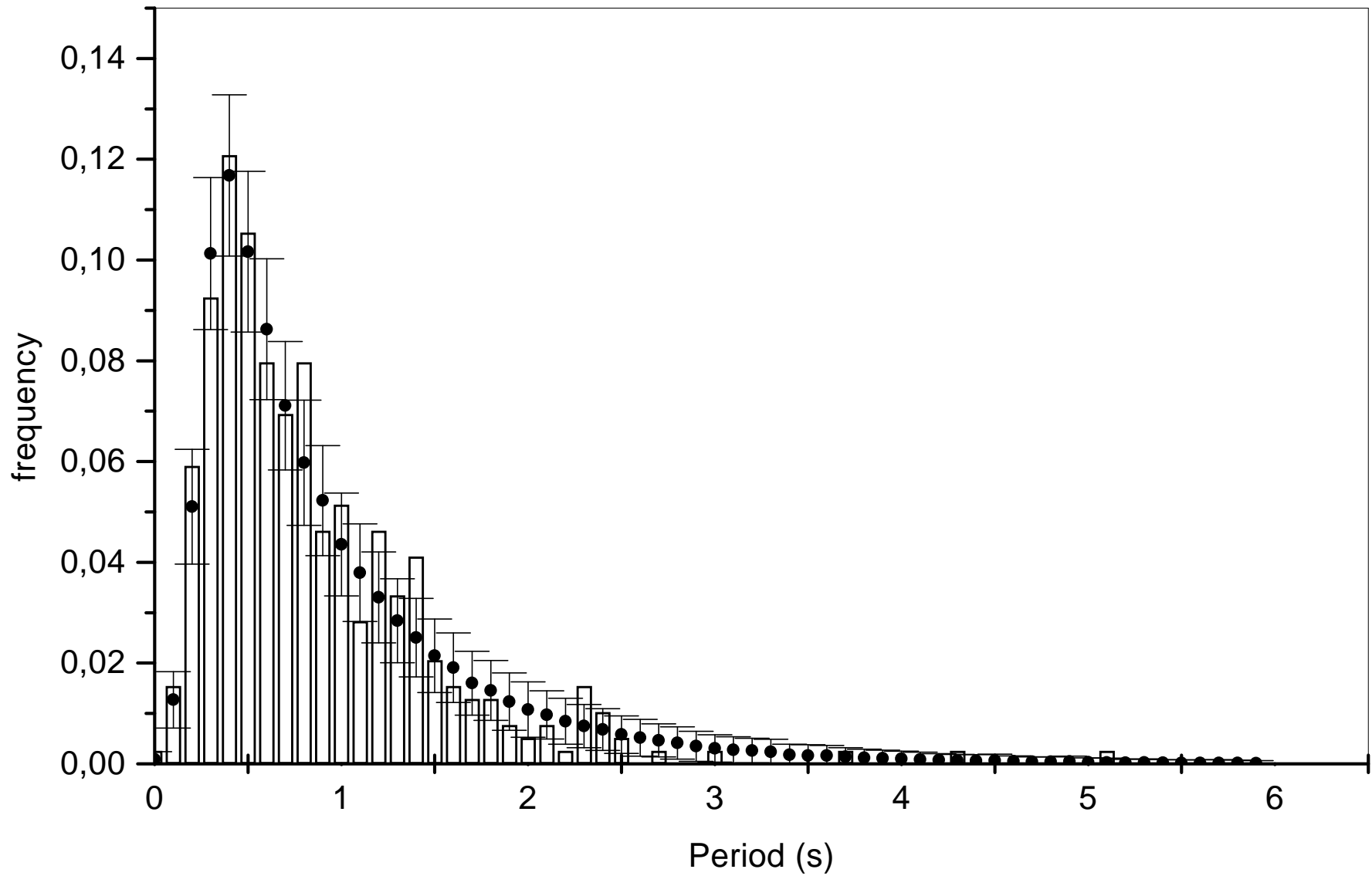
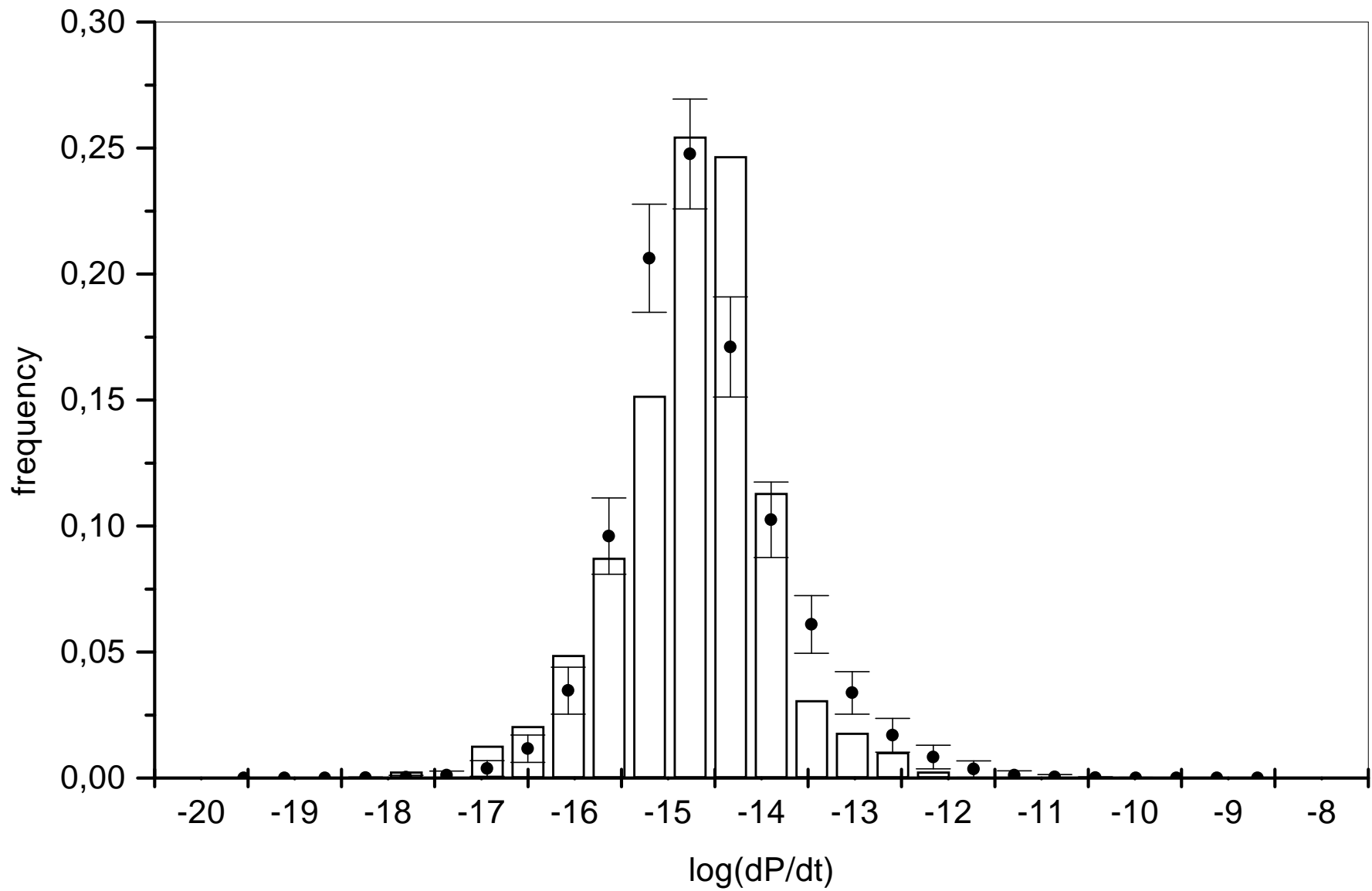
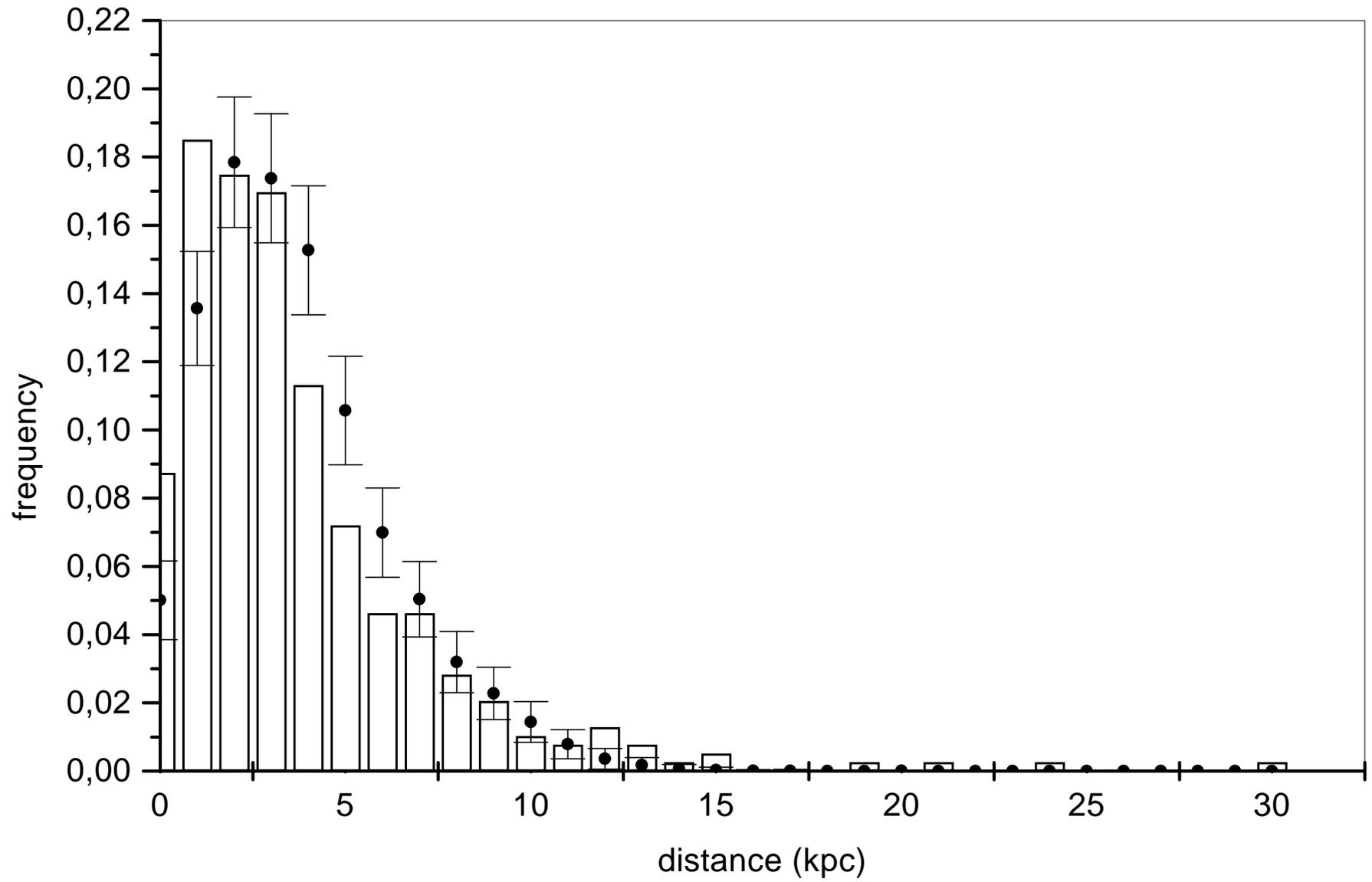
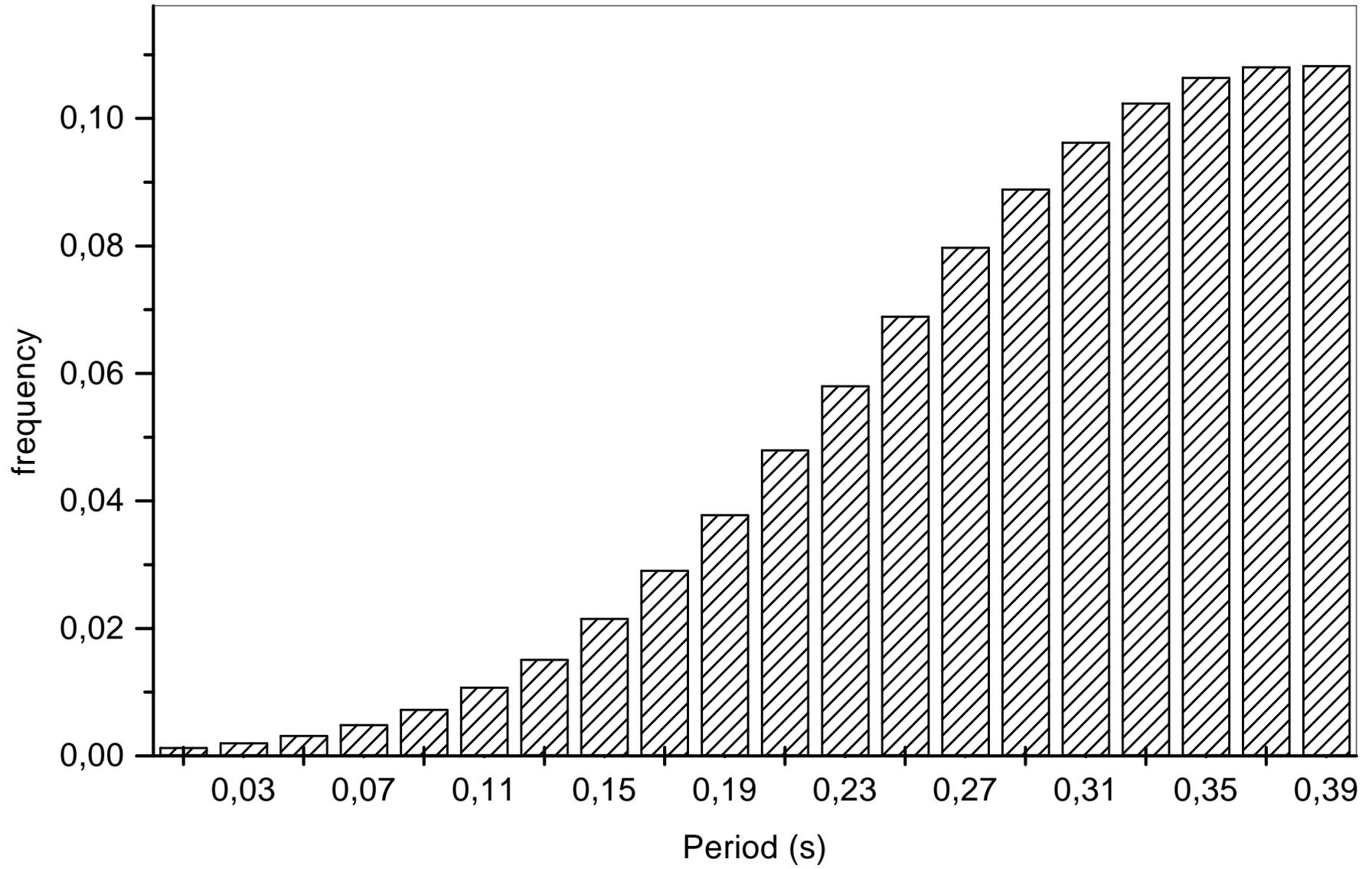
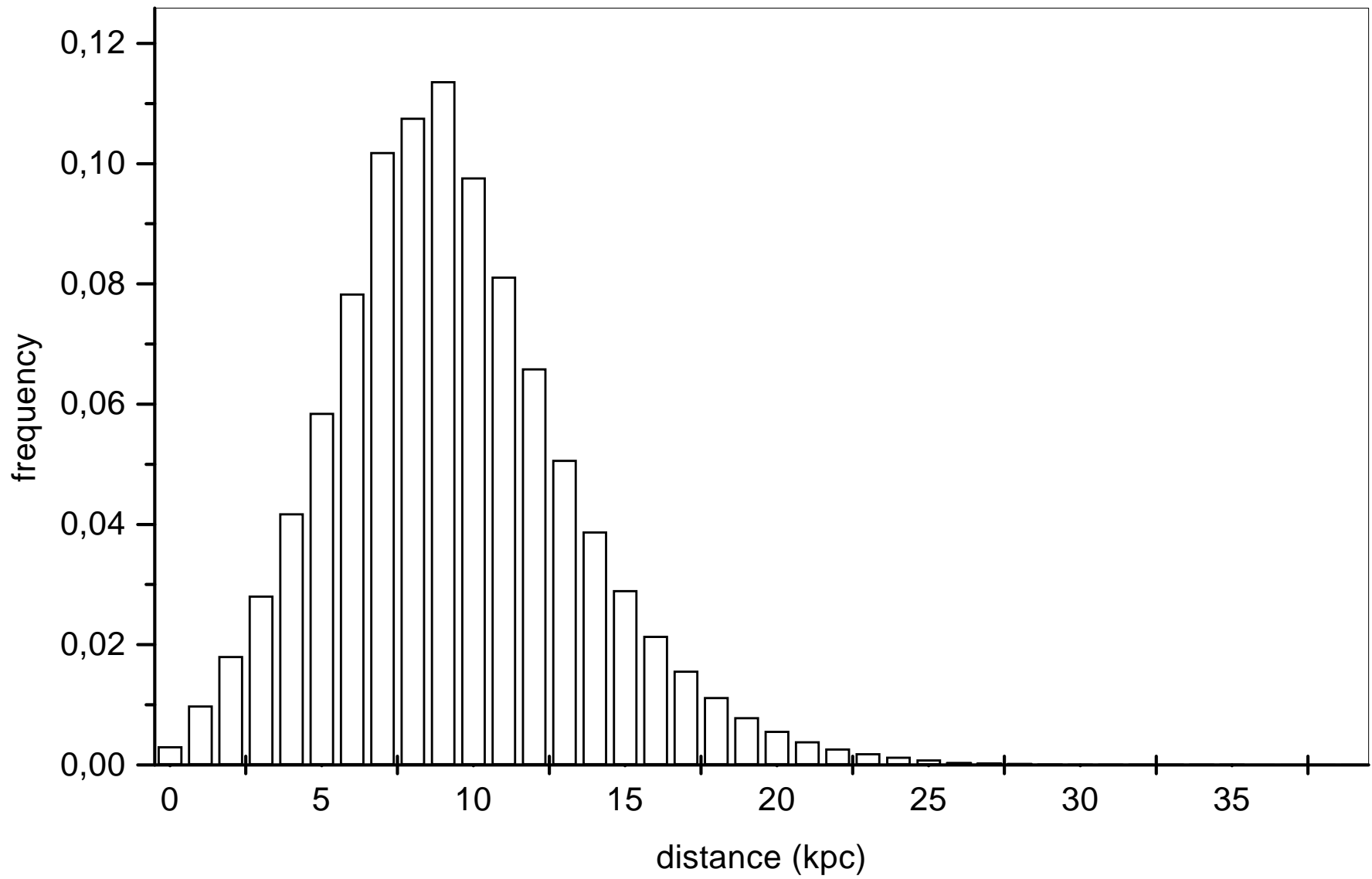
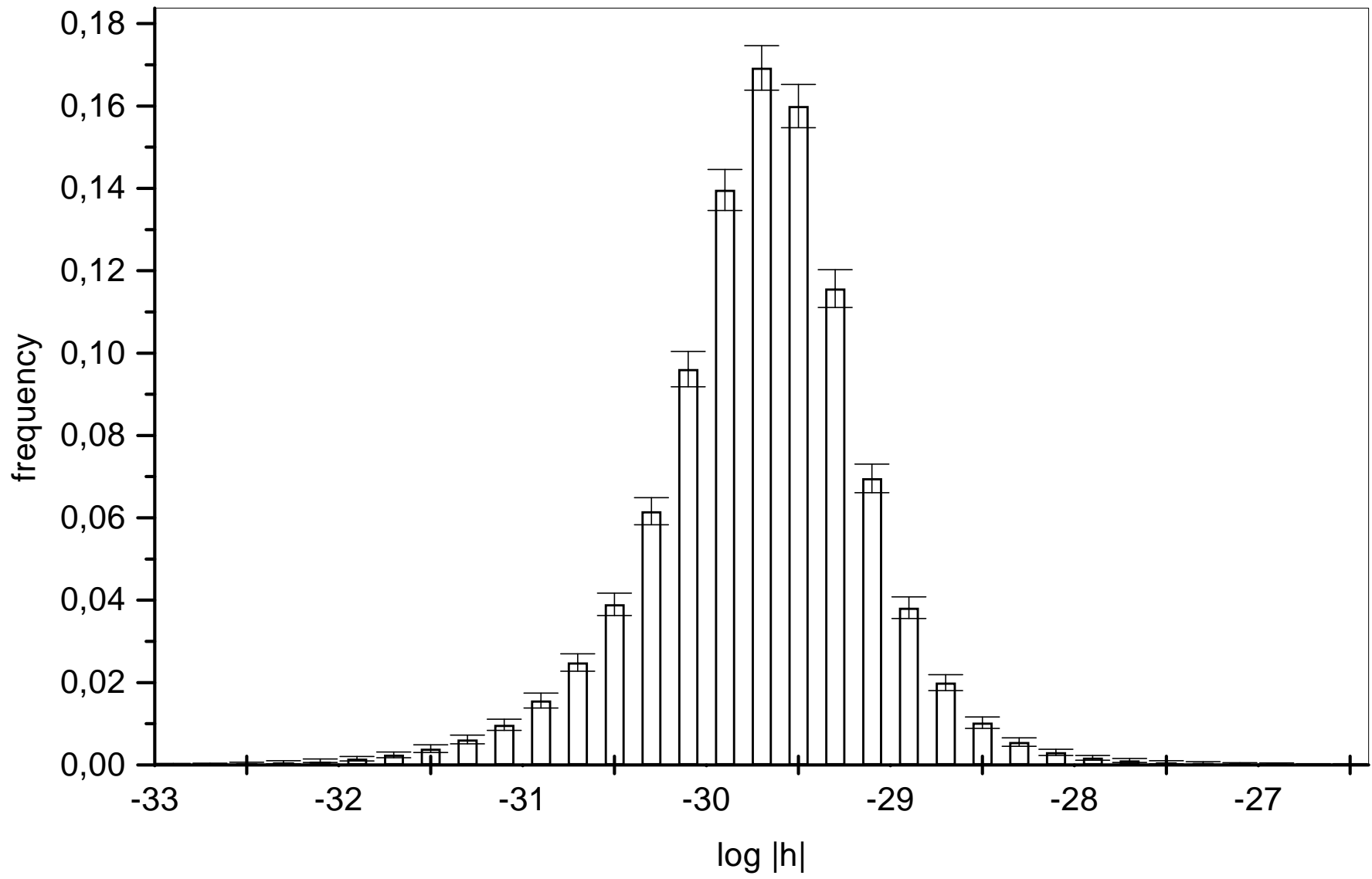
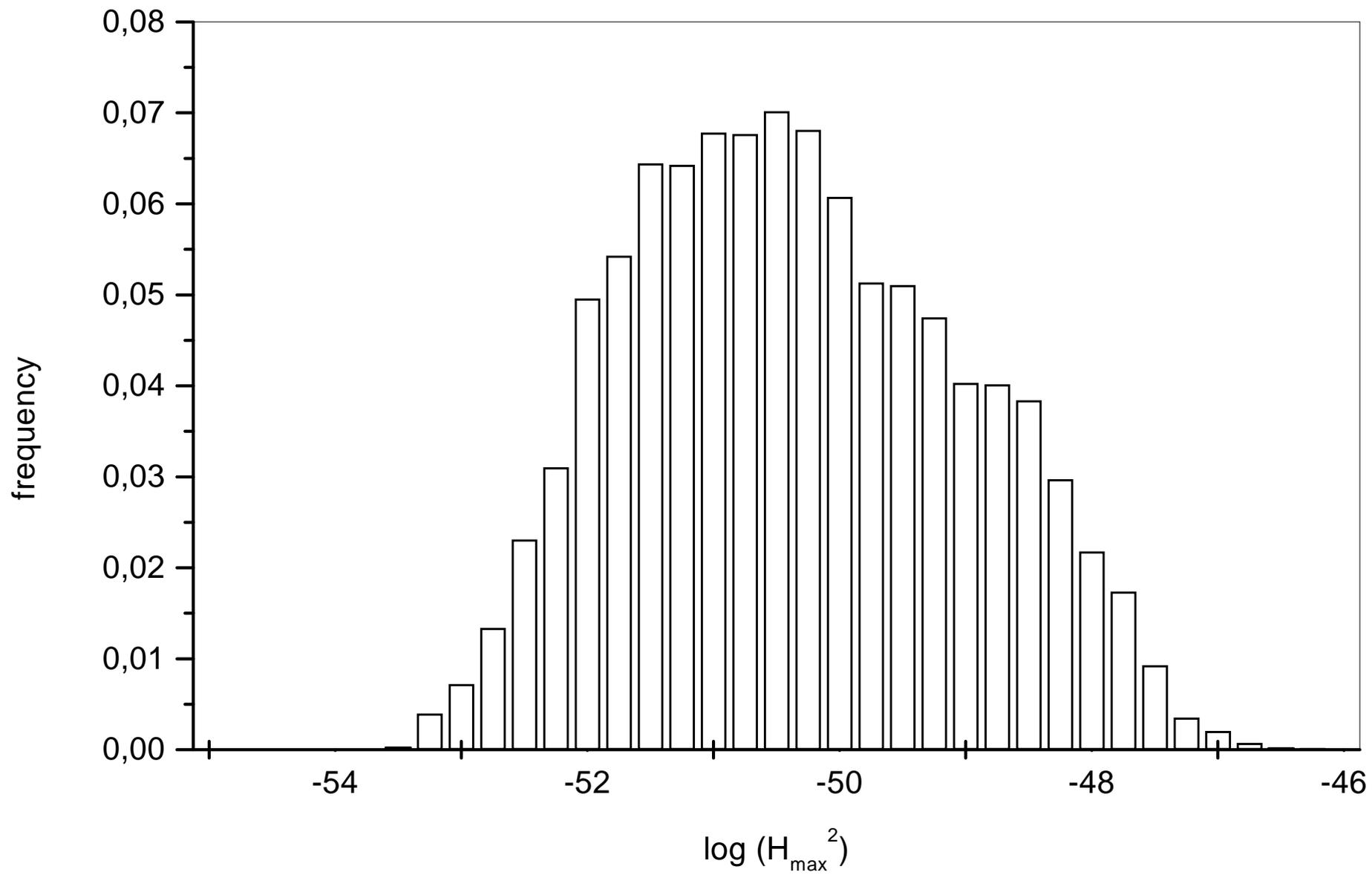
\end{document}